\documentclass[sn-mathphys,Numbered]{sn-jnl}

\usepackage{graphicx}%
\usepackage{multirow}%
\usepackage{amsmath,amssymb,amsfonts}%
\usepackage{amsthm}%
\usepackage{mathrsfs}%
\usepackage[title]{appendix}%
\usepackage{xcolor}%
\usepackage{textcomp}%
\usepackage{manyfoot}%
\usepackage{booktabs}%
\usepackage{algorithm}%
\usepackage{algorithmicx}%
\usepackage{algpseudocode}%
\usepackage{listings}%
\usepackage{gensymb}

\begin{document}
\title[Article Title]{%
Temporal synthesis of optical nonlinearity through synergy of spectrally-tuneable electron and phonon dynamics in a metamaterial 
}


\author[1]{\fnm{Jingyi} \sur{Wu}} 

\author[1]{\fnm{Anton Yu.} \sur{Bykov}} 

\author[1]{\fnm{Anastasiia} \sur{Zaleska}} 

\author[1]{\fnm{Anatoly V.} \sur{Zayats}} 

\affil[1]{\orgdiv{Department of Physics and London Centre for Nanotechnology}, \orgname{King's College London}, \orgaddress{\street{Strand}, \city{London}, \postcode{WC2R 2LS}, \country{UK}}}

\abstract{Manipulating intensity, phase and polarization of the electromagnetic fields on ultrafast timescales is essential for all-optical switching, optical information processing and development of novel time-variant media. Noble metal based plasmonics has provided numerous platforms for optical switching and control, enabled by strong local field enhancement, artificially engineered dispersion and strong Kerr-type free-electron nonlinearities. However, precise control over switching times and spectrum remains challenging, commonly limited by the relaxation of hot-electron gas on picosecond time scales and the band structure of materials. Here we experimentally demonstrate the strong and tuneable nonlinearity in a metamaterial on a mirror geometry, controlled by the wavelength of excitation, which imprints a specific non-uniform hot-electron population distribution and drives targeted electron and lattice dynamics. The interplay of electromagnetic, electronic and mechanical energy exchange allows us to achieve sub-300~fs timescales in the recovery of optical constants in the selected spectral domains, where the modulation surpasses the limitations imposed by the inherent material response of metamaterial components, owing to emergence of a Fano-type destructive interference with acoustic vibrations of the metamaterial, featured in reflection but not in transmission. The observed effects are highly spectrally selective and sensitive to the polarisation properties of light and the Fabry-Perot modes of the metamaterial, opening a pathway for controlling the switching rates by spectral selection and nanostructure design. The capability to manipulate temporal, spectral and mechanical aspects of light-matter interactions underscores new potential nonlinear applications where polarisation diversity, spectral selectivity and fast modulation are important.} 


\maketitle

\section{Introduction}\label{secIntr}

Ultrafast optical nonlinearities are highly desirable for numerous applications in laser physics, integrated photonics, active nanophotonics and quantum technologies \cite{Kaur2012,Koya2023}. Dielectric and plasmonic nanostructures and metamaterials are widely used for enhancing nonlinear optical processes. While the nonlinear efficiency is readily controlled by the field enhancement, the temporal response is more difficult to manipulate as it is related to the intrinsic properties of material constituents.

Electronic nonlinearities in plasmonic nanostructures are related to rapid heating of electron gas, which results in the Kerr-type nonlinear response, and provide many opportunities to design the subwavelength-size active photonic devices for enhanced and ultrafast light-matter interaction \cite{Khur2023}. Impulsive heating of the electron gas initiates a spectrum of dynamic phenomena and thermal effects in nanostructures that enable ultrafast control over optical waveforms through the nonlinearity of hot-electron gas
\cite{Kra2018,Khur2023}. The modulation rate in this case is limited by characteristic lifetimes of hot-electron population in metal \cite{Fatti2000,Arbouet2003}, however nanostructuring was shown to impact the response time by engineering the hot-electron diffusion away from the interfaces \cite{Nic2019}. 

Mechanical degrees of freedom can also be exploited related to the energy transfer from the hot-electron ensemble to mechanical motion of nanostructures, resulting in the coherent excitation of acoustic phonons \cite{Temnov2012,Frischwasser2022}. This phenomenon has been observed in both individual nanoparticles \cite{Pelton2009,Bykov2021} and complex nanostructures \cite{Ruello2015,LeGuyader_2008,Bragas2023}. These excitations can be selectively manipulated through temporal or spectral tailoring of the impinging light \cite{YuOuyang2018} and may interfere with the transient response driven by the Kerr-type nonlinearity of hot electrons. Such interactions provide a distinct and independent mechanism to control optical properties of nanostructures, and are integral to enhancing the efficiency and effectiveness of the metamaterials in nanophotonic devices. 

Here, we demonstrate the strong and tuneable nonlinearity driven by spectrally-tuneable non-uniform hot-electron population and associated excitation of acoustic vibrations in a plasmonic metamaterial on a mirror geometry. Previously, strong ultrafast nonlinear optical modulation was demonstrated using plasmonic nanorod metamaterials in the epsilon-near-zero (ENZ) regime for the transmission of TM-polarized light \cite{Wurtz2011}. The spectral features of the metamaterial on a mirror geometry offer a set of reflection resonances for both TM- and TE- polarised light, allowing the polarization diversity of both signal and control light as well as exploitation of additional effects associated with the heat diffusion in the electron gas population and excitation of acoustic phonons, hidden in the transmission near the ENZ wavelength due to strong absorption. We experimentally demonstrate the spectral control of hot-carrier and lattice dynamics in a gold nanorod metamaterial and their interplay in the nonlinear light-matter interaction regime. At the Fabry-Perot reflection resonance of the studied structure, the enhancement of the modulation amplitude and unique temporal response of the metamaterial is observed. The strong electron temperature increase in a mirror to which the metamaterial is connected may introduce a backward-hot electron diffusion, which is usually not considered, and impulsive excitation of phonons in metamaterial. The interplay between mechanical and electronic degrees of freedom enables ultrafast modulation of optical reflection on a sub-300~fs timescale, beyond the constraints imposed by the material response of the metamaterial components, which further can be controlled with the choice of the excitation wavelength. At the same time, transmission of the metamaterial is largely unaffected. We have further identified several acoustic vibrational modes of both the individual nanorods and the metamaterial slab as a whole
The dynamic interaction between acoustic and electronic modes of the metamaterial allows us to control the response type of the nonlinearity beyond the characteristic timescales of plasmonic hot-electron relaxation alone. 

\section{Results}

\begin{figure}[t!]%
	\centering
	\includegraphics[width=1.\textwidth]{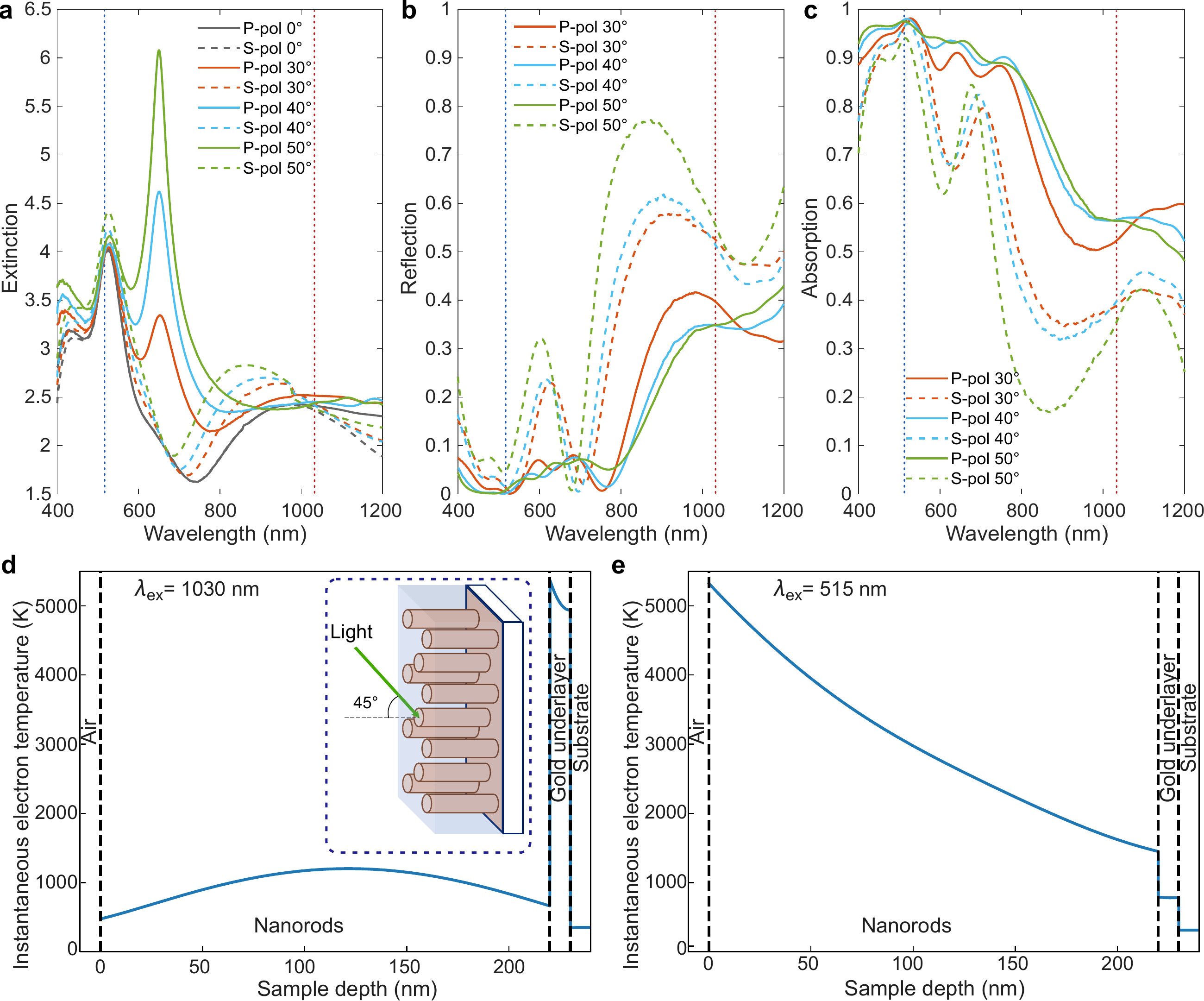}
	\caption[Optical properties of the metamaterial on a mirror system.]{\textbf{Optical properties of the metamaterial on a mirror system.} (a) Extinction, (b) reflection, and (c) absorption of the nanorod metamaterial for various polarisations and angles of incidence. The metamaterial parameters are: nanorod radius 14~nm, length 220~nm and spacing $\sim$70~nm. (d,e) Instantaneous hot-electron temperature distributions at different excitation wavelengths: (d) 1030~nm and (e) 515~nm, normalised to the same absorbed energy density 1 mJ cm$^{-2}$.  
 }\label{figERA}
\end{figure}

We explored the interplay between hot-electron dynamics and acoustic phonons in the metamaterial based on an array of plasmonic (gold) nanorods (length 220~nm, diameter 28~nm  and spacing 70~nm) embedded in a dielectric matrix of fused alumina. The metamaterial is grown on a glass substrate and has a 8-nm-thick gold underlayer (Fig.~1 insert; the details of the fabrication process are described in Methods). The nanorod metamaterial behaves as an optically anisotropic medium and, for the extraordinary wave, exhibits the ENZ condition at a wavelength of 650~nm, where the real part of the effective permittivity $\Re\{\epsilon_z\}$ approaches zero. This corresponds to the transition between the elliptic and hyperbolic dispersion regimes of the metamaterial \cite{Poddubny2013,DJRoth2024}. Around this wavelength, a strong extinction peak is observed under oblique incidence with $p$-polarised illumination (Fig.~\ref{figERA}a). The reflection spectra (Fig.~\ref{figERA}b), on the other hand, exhibit a series of pronounced resonances in the hyperbolic regime ($\lambda > \lambda_{ENZ}$), characteristic of leaky waveguided modes within the metamaterial layer (Fabry-Perot type resonances), which effectively modulate the absorption spectra (Fig.~\ref{figERA}c). 
The transient nonlinear optical properties around the ENZ wavelength are strongly enhanced~\cite{Reshef2019,Neira2015}. Beyond this wavelength, the metamaterial exhibits weaker absorption, enabling detailed analysis of the transient processes associated with the guided modes \cite{Vasilantonakis2015}, which can be observed in reflection for both polarizations of light.

\subsection{Spectrally tuneable hot-electron population.}

\begin{figure}[t!]
	\begin{center}
	\includegraphics[width=1.0\textwidth]{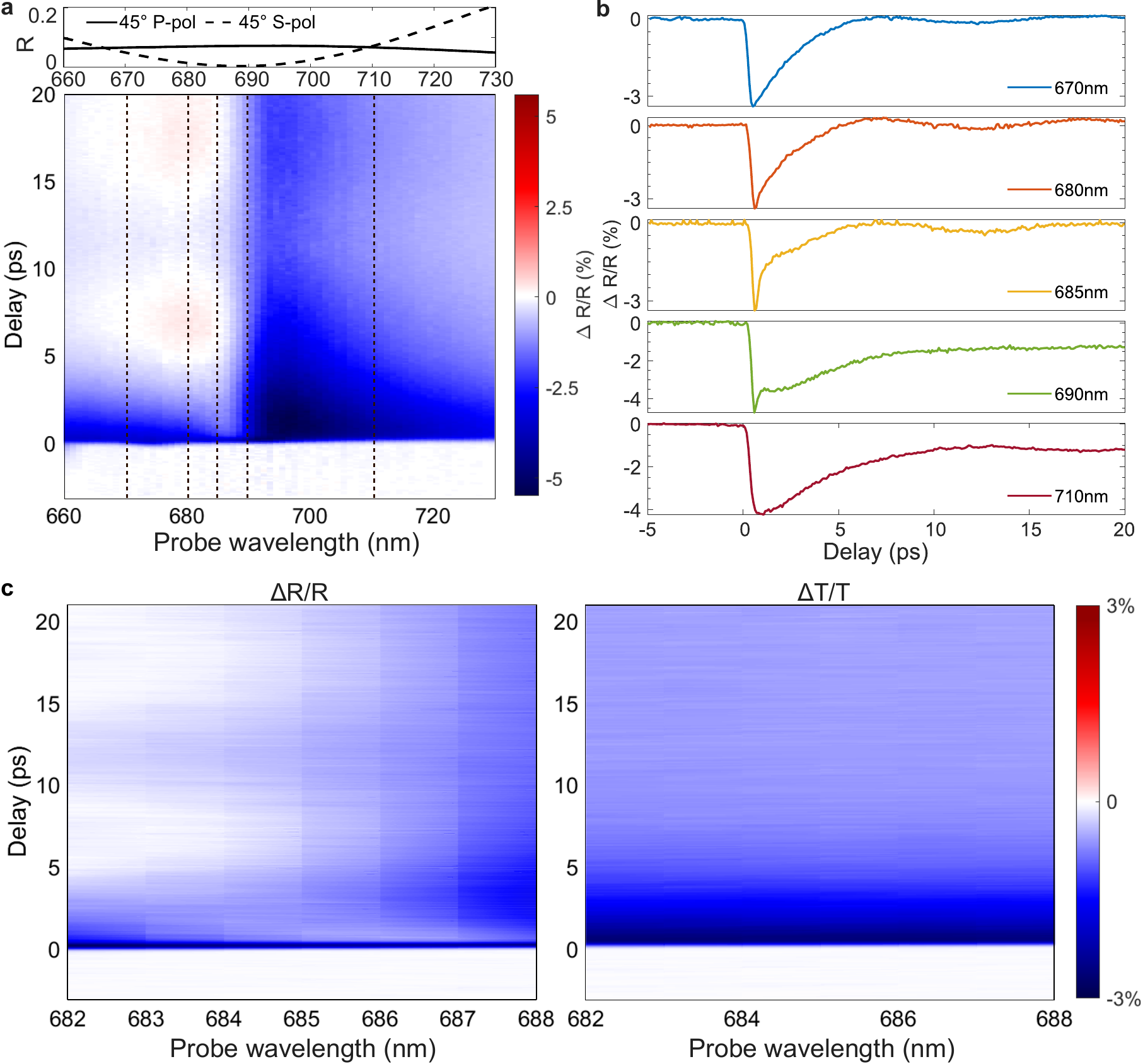}
	\caption[Transient optical spectra of the metamaterial under the 1030~nm excitation.]{\textbf{Transient optical spectra of the metamaterial under the 1030~nm excitation.} (a) Top: reflection spectra of the metamaterial on a mirror in the ground state measured at 45$\circ$ angle of incidence. Bottom: transient reflection spectra measured at the 1030~nm excitation. (b) Time dependent reflectivity at selected wavelengths of 670~nm, 680~nm, 685~nm, 690~nm and 710~nm (the corresponding cross-sections in (a) are indicated with dashed lines). (c) High-resolution transient spectra of (left) reflection and (right) transmission measured simultaneously.}\label{fig3Dwith3}
    \end{center} 
\end{figure}

It is known that the excitation of optically-thick metamaterials produces the non-uniform temperature distributions within the structure that can be tailored by the excitation wavelength \cite{Nic2019,Schirato2020}. Light absorption governed by the electromagnetic mode structure of the metamaterial placed onto a gold mirror results in the increase of the electron temperature in both the nanorods and the mirror with a strong wavelength-dependent spatial profile (Fig.~\ref{figERA}d,e; see Methods for the details of the simulations). Absolute values of the temperature rise are similar for the same powers of the absorbed energy at different wavelengths. For the excitation in the NIR spectral range (1030~nm), the gold mirror absorbs the most of optical energy and the electron temperature in the mirror is changed dramatically while the electron temperature in the metamaterial almost five times smaller with small variations along the nanorods. The situation is opposite for the excitation in the visible spectral range (515~nm) with the strongest temperature increase observed in the nanorods and five times smaller in the gold mirror. Strong electron temperature gradient is present along the nanorods in this case with the maximum temperature increase at the illuminated interface. 

The reflection modulation of nearly 10\% under the moderate NIR excitation intensities is observed due to the increased reflection upon hot-electron generation (Fig.~\ref{fig3Dwith3}a). The strongest change of reflection is observed in the vicinity of the guided mode at a wavelength of 690~nm. Surprisingly, the transient signal with opposite signs on different sides of the resonance, typically expected in the case of the shift of the resonant mode
is not observed. At the shorter wavelengths, a pronounced asymmetric Fano-type spectral shape is present, and this asymmetry manifests itself in the time domain as an emergence of very short initial decay observed in this spectral range, e.g., at wavelengths of 685~nm and 690~nm (Fig.~\ref{fig3Dwith3}b). The observed Fano-type interference in the time domain demonstrates strong spectral selectivity and sensitivity and substantial modifications of the relaxation time of the transient reflection across the spectral domain. A wavelength-independent oscillatory behavior is also observed with the minima at approximately 7 and 19~ps post-excitation (after time-zero in Fig. \ref{fig3Dwith3}b). Notably, no such effects are observed in the transient transmission of the metamaterial, and overall values of modulation amplitude in transmission remains lower despite the whole optical thickness of the slab and the mirror playing a role in the response (Fig.~\ref{fig3Dwith3}c). The transient transmission decays monotonically in several picoseconds and is spectrally featureless. No oscillatory behaviour is observed in transmission, in stark contrast with the dynamics of the reflectance.  

 We suggest it is the interaction of the probe light with the hot-electron gas in the mirror that contributes to the overall increase of the reflection, producing the observed fast relaxation dynamics through destructive interference of multiple reflections in a multilayer system. The thermal expansion of the underlayer could also serve as the nanoscale transducer for the excitation of the rich spectrum of acoustic vibrations in the metamaterial, as will be discussed below. The strong electron temperature increase in a mirror to which the metamaterial is connected introduces a backward-hot electron diffusion, which is usually not considered in the majority of the studies, despite similar geometries being routinely present in nanoscale structures due to technological steps of substrate preparation. The exact role of this process and the peculiarities of hot-carrier dynamics at the interface between the mirror and the nanorods depends on the values of interface thermal conductivity and electron diffusion between the mirror and the base of the nanorods, evaluation of which is challenging in real materials, but which is important mechanism demonstrated in this study and should not be overlooked.

\begin{figure}[t!]%
	\begin{center}
	\includegraphics[width=0.5\textwidth]{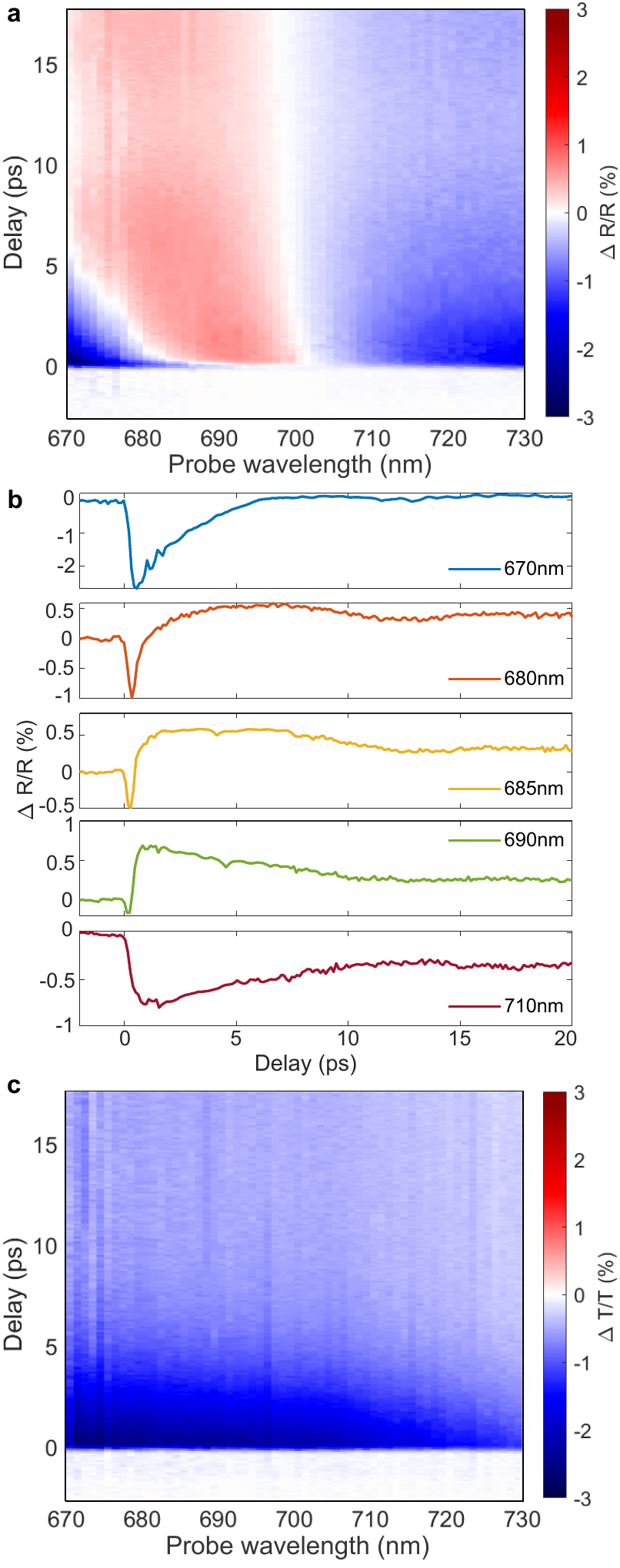}
	\caption[Transient optical spectra of the metamaterial under the 515~nm excitation.]{\textbf{Transient optical spectra of the metamaterial under the 515~nm excitation.} Transient (a,b) reflection and (c) transmission spectra measured with the excitation at a wavelength of 515~nm and a power of 10~mW, which is 5 times smaller than that of the IR pump in Fig.~\ref{fig3Dwith3}.}
 \label{figGreenPump}
    \end{center}
\end{figure}

The situation is drastically different for a visible light excitation which reduces electron heating in the mirror (Fig.~\ref{figGreenPump}). While the transient transmission mainly follows the same trend as observed with the NIR excitation, the reflection signal now undergoes a typical sign-changing behaviour, indicating the shift and broadening of a Fabry-Perot resonance of the metamaterial slab. The stronger electron heating, which in this case primarily takes place in the nanorods (Fig.~\ref{figERA}e), results in the stronger modulation amplitude, compared to the NIR excitation: similar modulation amplitude is observed with 5 times smaller excitation intensity (Fig.~\ref{figGreenPump}). Overall, the transient signals beyond 20\% in reflection have been observed with twice less power than under the NIR excition. Interestingly, the electron temperature increase localised primarily in the nanorods also mitigates the optical damage to the sample, which was observed to be limited by the temperature increase in the gold mirror. 

\begin{figure}[ht]
    \centering
    \includegraphics[width=1.0\textwidth]{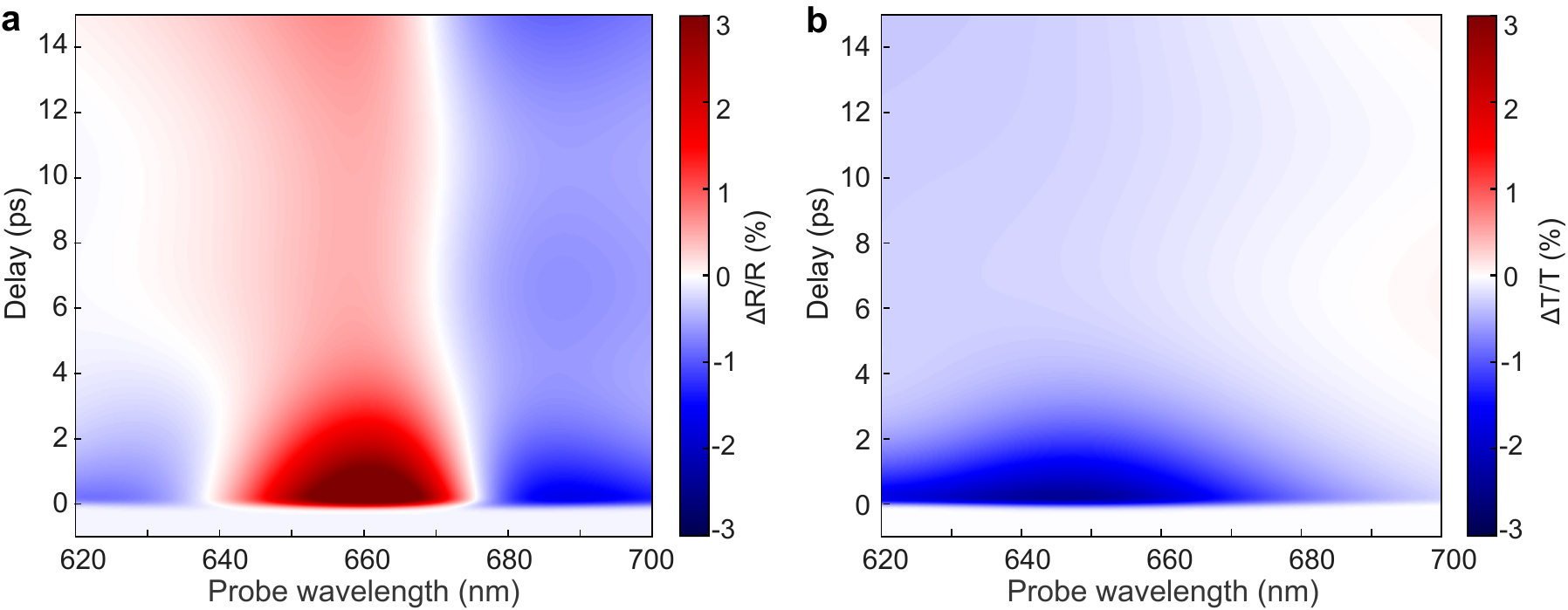}
    \caption[Numerical simulations of the transient response.]{\textbf{Numerical simulations of the transient response.} Simulated transient (a) reflection and (b) transmission spectra of the metamaterial on a mirror system under the excitation at a wavelength of 515~nm.}
    \label{figTTM_dRdT}
\end{figure}

The numerical modelling (see details in Methods) reproduces the overall transient behaviour in both transmission and reflection (Fig.~\ref{figTTM_dRdT}a-b). We would like to emphasize that despite the success in modelling the response of nanorod composite under visible illumination, the model faces challenges to replicate the most important features of the NIR-induced response, highlighting yet again the importance of detailed treatment of hot-carrier diffusion within optical metamaterials and between the nanorods and the mirror.

\subsection{Acoustic phonon properties.}

\begin{figure}[ht]%
	\begin{center}
	\includegraphics[width=0.85\textwidth]{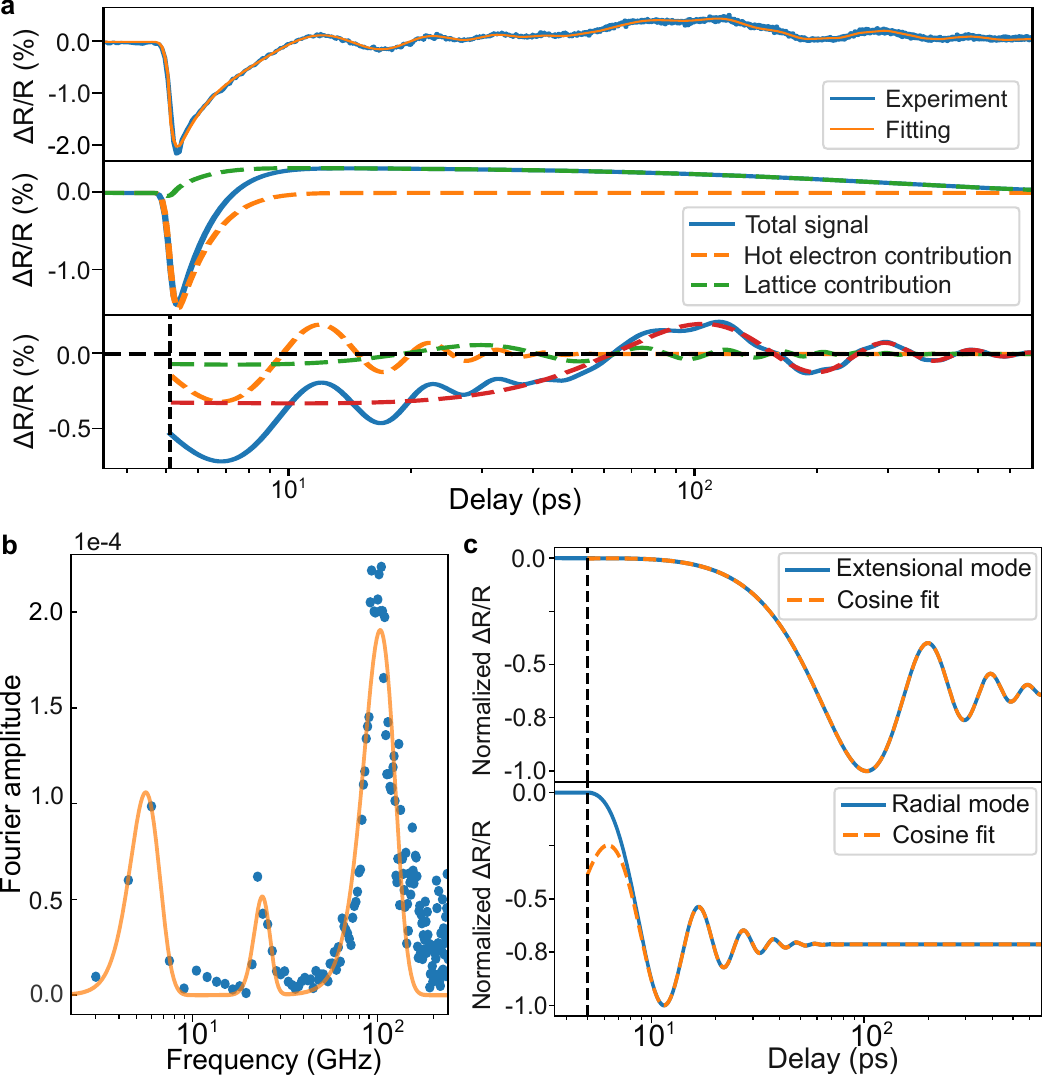} %
	\caption[Spectrum of acoustic phonons.]{\textbf{Spectrum of acoustic phonons.} (a) Transient reflection trace at a wavelength of 680~nm under the 1030~nm excitation, with corresponding model fit (upper panel), separated into aperiodic (middle panel) and periodic (lower panel) parts. Log scale used for the horizontal axis for better display of features with different temporal scales. (b) Spectrum of the periodic part of (a) obtained with fast Fourier transform (orange lines are Gaussian fit). (c) Solutions of Eq.~\ref{vibrationModel} for (top) extensional mode of the nanorod and (bottom) radial breathing mode of the nanorod. Dashed orange lines represent cosine fits and demonstrate different recovered initial phases}\label{figVibFit}
    \end{center}
\end{figure}

The long-lived weakly wavelength-dependent oscillations in the transient response can be attributed to the excitation of various acoustic modes within the metamaterials (Fig.~\ref{fig3Dwith3}). Over the time scale of 500~ps, multiple superimposed periodic signals are observed featuring periods ranging from a few picoseconds to 100s~ps (Fig.~\ref{figVibFit}a). A detailed analysis of the frequencies of the modes done by the fast Fourier transform and least-square fitting (Fig.~\ref{figVibFit}b) identified three damped oscillations with the frequencies of 5 GHz, 23 GHz and 97 GHz and corresponding decay times of 190, 120, and 10 ps, respectively. 
We can identify the vibrations with the lowest and highest frequencies as fundamental breathing radial and extensional modes of a weakly coupled individual nanorods in the metamaterial \cite{Yu2013}:
\begin{eqnarray}
\nu_{br}&=&\frac{\phi_0 c}{2\pi R},\nonumber \\
\nu_{ext}&=&\frac{1}{2L}\sqrt{\frac{E}{\rho}},
\end{eqnarray}
where $R$ and $L$ are the radius and length of the nanorod, $c$, $E$, and $\rho$ are the longitudinal speed of sound, elastic modulus and density of gold, respectively, and the eigenvalue $\phi_0$ is the first root of the equation $\phi_n J_0(\phi_n) = (1 - 2 \sigma)J_1(\phi_n)/(1 - \sigma)$, with $\sigma$ being the Poisson ratio, roughly equal to 2.28. Taking into consideration the characteristic dimensions of the metamaterial, this provides frequency estimates of 4.6 and 85 GHz, respectively, that match the experimental observations. The slight deviations of the calculated frequencies from the experimental ones are probably related to deviation of the elastic modulus of the electrodeposited gold from the tabulated values of bulk material, and/or small systematic errors in the determination of radius and length of nanorods from the SEM data. Since in the experiment, the control and probe beams illuminate multiple nanorods, the decay times of these vibrations are mainly a signature of inhomogeneous broadening rather than intrinsic decay times which have been shown to be of the order of 100 and 400 ps for the breathing and extensional vibrations of isolated nanorods in air \cite{Yu2013} (the alumina matrix may also contribute to the broadening but on a smaller scale than the inhomogeneous damping). The mode at the frequency of 23 GHz is consistent with the standing longitudinal wave in the alumina matrix (estimated frequency 22.7 GHz), with the period corresponding to the round trip across the film. 

Two low frequency vibrational modes are excited as a roughly cosine function with initial phases close to 20$^\circ$ as determined from the fit (Fig.~\ref{figVibFit}a), consistent with previous observations \cite{Wang2007,WangJ2007,Perner2000}, signifying that the initial thermal expansion, that depends on the dynamics of electron and lattice temperatures happens on the timescale much faster than the period of the corresponding oscillation. The radial breathing mode of the nanorods, however possesses an initial phase closer to 70$^\circ$. 
The latter could be understood by considering the simplistic dynamic model \cite{Perner2000}:
\begin{eqnarray}
\label{vibrationModel}
\frac{d^2x}{dt^2} &+& \frac{2}{\tau}\frac{dx}{dt} + \omega_0^2x = \frac{A}{m}\sigma(t), \nonumber \\
\sigma(t) &=& \gamma C_l\Delta T_l(t) - \frac{2}{3}\Delta E_e(t),
\end{eqnarray}
where $x$ is a vibration coordinate, 
$\omega_0, \tau$ are the fundamental frequency and decay constant, that can be taken from fitting of experimental data, $A$ and $m$ are the area and mass of nanorod, respectively, and $\sigma(t)$ is the transient stress, created by the rise and decay of hot electron and lattice temperatures. The latter contains two terms \cite{PhysRevB.49.15046, GUSEV199276}: one caused by the lattice anharmonism proportional to the temperature of the lattice ($C_l$ is the heat capacity of the lattice, $\gamma \approx -2.7$ is the Grüneisen constant \cite{GUSEV199276}), and another caused by the hot electron pressure, proportional to the total energy $\Delta E_e$ stored in the hot-electron bath.

Solving Eq.~\ref{vibrationModel} allows to determine the initial phases for extensional and radial vibrations of a gold nanorod, which are found to be around 12$^\circ$, 52$^\circ$, respectively, in a close agreement with the experimental observations (Fig. \ref{figVibFit}c). We do not apply this model for analysis of the standing acoustic wave in the composite, since it is not directly excited by the hot-electron dynamics, but rather through mechanical coupling with the other parts of the structure. 

We would like to emphasize that such temporal shaping of transient response, observed under the NIR excitation occurs only in reflection, where the optical resonance associated with the guided mode in the hyperbolic metamaterial composite is present, providing the necessary phase flip. In transmission, on the other hand, the optical modulation produces the changes in optical constants decaying much slower--on the order of few picoseconds--consistent with the hot-electron decay in gold (Figs.~\ref{fig3Dwith3}c,~\ref{figGreenPump}b). Notably, our transient spectra during long delays also provided distinct evidence for the superposition of diverse excitation and resonances within gold nanorod metamaterial induced by ultrafast pulses.

\section{Discussion}
We used a plasmonic nanorod metamaterial on a mirror to demonstrate and control ultrafast all-optical processes and excitation of acoustic vibrational modes. The nonlinear optical response of the plasmonic nanorod metamaterial was previously demonstrated in transmission only for TM-polarised signal light at around the ENZ wavelength where it is limited by strong absorption of the metamaterial at the operational wavelength. The rich optical reflection spectrum of the metamaterial offers alternative optical resonances within the hyperbolic dispersion range, corresponding to the set of both TE and TM guided modes intrinsic to the structure \cite{Vasilantonakis2015}, which allows the polarisation diversity of the signal and control light. The differences in absorption magnitude and absorption profiles along the nanorods at different wavelengths result in a stronger sensitivity of the reflection to the acoustic vibrations and a plasmonic mirror effects, which are negligible in transmission but important at low absorption conditions in reflection. The nonlinear response exhibits the timescales shorter than the characteristic hot-carrier relaxation periods in the materials constituting the metamaterials due to emergence of a Fano-type destructive interference with acoustic vibrations of the metamaterial.   

The strong, excitation-wavelength-dependent electron temperature increase in a mirror to which the metamaterial is connected introduces a backward-hot electron diffusion, important for both temporal characteristics of nonlinearity and controls the excitation of vibrational acoustic modes. At the same time, the damage threshold is significantly reduced at the wavelengths for which absorption in a mirror dominates.  

The acoustic vibrations of the constituting elements of the metamaterial excited by the decay of hot carriers behave as a superposition of sine or cosine functions in time trajectory \cite{Perner2000,Thomsen1986}. Such acoustic excitations have been extensively studied in both isolated nanoparticles in various environments \cite{Deacon2017,Bragas2023}, where they could be used as nanoscale probes of local elasticity of materials and local viscosity of surrounding medium, and in composite nanostructures where the design of the structure offers wavelength and polarisation sensitivity \cite{Frenzel2019,Jansen2023} and as ultra-compact sources of surface acoustic waves. 

The observed interplay between hot-electron and mechanical dynamics of the metamaterial provides new insights in the understanding of light-matter interactions in complex nanostructured materials and expands the potential for designing nonlinear optical properties for nanophotonic and acousto-optic applications, time-varying optics, and nanoscale metrology with acoustic vibrations.

\section{Materials and Methods}\label{secMethod}

\subsection{Sample fabrication}\label{subsecFab}

\begin{figure}[ht]
     \centering
     \includegraphics[width = 0.9\textwidth]{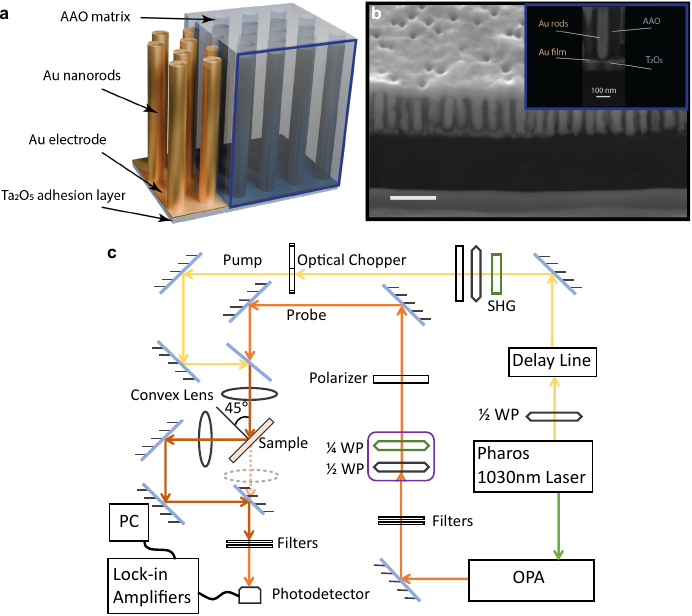}
     \caption[Metamaterial structure and the experimental setup.]{\textbf{The metamaterial structure and the experimental setup.}(a) Schematic and (b) the SEM image of the cross-sectional view of the metamaterial (scale bar is 200~nm); insert is the cross-section image of a single nanorod. 
     (c) Schematic diagram of the pump-probe spectroscopy setup: (purple box) the group of waveplates for adjustment of the input polarisation states; (green box) a removable SHG module for generation of a 515~nm light.}
     \label{figSEM} 
 \end{figure}

The plasmonic anisotropic metamaterial based on an array of gold nanorods (Fig.~\ref{figSEM}a,b) was fabricated using the alumina template method \cite{DJRoth2024,Zaleska2024}. Thin aluminium films of up to 400~nm thickness were first anodised in 0.3~M sulfuric acid at 25~V to create a porous anodised aluminium oxide (AAO) template. After removing the barrier layer with a 30~mM NaOH etching solution, gold nanorods were electrochemically grown within a substrate-supported alumina template, which included a sputtered 10~nm Ta$_2$O$_5$ adhesion layer and an 8~nm Au film acting as the working electrode for electrochemical deposition. The diameter and periodicity of the Au nanorods were determined by the geometry of the AAO templates and were controlled by anodisation conditions such as the choice of electrolyte, its temperature, and the anodisation voltage. The length of the nanorod structures was controlled by the duration of the electrochemical deposition. 

\subsection{Time-resolved measurements}\label{PumpProbe}

The time-resolved transient measurements were performed using an optical parametric amplifier (OPA) to provide a tuneable probe beam with the pulse duration of 150 fs. The OPA is pumped by an Yb-~amplified laser system (Light Conversion Pharos), which also provided a pump beam at a wavelength of 1030~nm  and $\approx$ 250~fs pulse duration (Fig.~\ref{figSEM}c). The output of the pump laser was frequency doubled in an additional Barium Borate (BBO) crystal to realise the excitation at the wavelength of 515 nm. The TM-polarised excitation pulses and TE-polarised probe pulses were incident on the sample at an angle of 45$^o$. The experiment is designed to allow measurements of both transient reflection and transmission. 

\subsection{Modelling electron dynamics}\label{subsecTTM}

To model the electron and lattice dynamics in the metamaterial, we employed the two temperature model (TTM) to describe the equilibrium hot-electron population, established by initial thermalisation through ultrafast electron-electron collisions, that subsequently transfers energy to the crystal lattice and the environment through electron-phonon and phonon-phonon scattering \cite{Peruch2017, WangJ2007}. Due to the lateral cross-section of an individual nanorod ($\approx$28~nm) being much smaller than sizes of the pump and probe beams and a skin-depth in gold at both the pump and probe wavelengths, we ignore the in-plane variation of electron and lattice temperatures. This leads to the one-dimensional model, that without the loss of generality only needs to consider the heat transfer between the thermodynamic ensembles and 1D heat diffusion: 
 \begin{align}
     \label{EqTTM_e}
     C_e(T_e)\frac{\partial T_e}{\partial t} &= \nabla (k_e \nabla T_e) - G(T_e - T_l) + S(t,z),\\ 
     C_l\frac{\partial T_l}{\partial t} &= G(T_e - T_l)
     \label{EqTTM_l}
 \end{align}
 where $C_e$ and $C_l$ are the heat capacities of the electron and lattice subsystems respectively, $k_e$, $G$ and $S(t,z)$ are the electron thermal conductivity, the electron-phonon coupling factor, and the source term describing electron heating by the laser pulse, as a function of the time \cite{Chen2006307}. 

 At hot-electron temperatures much smaller than the Fermi temperature, $C_e$ can be approximated by a linear function $C_e = C_{e0} T_e$, with the proportionality constant $C_{e0} = \pi ^2 N k_B/2T_F$, following from the Sommerfeld theory of metals, where $N$ is the number density of atoms, $k_B$ is the Boltzmann constant, $T_F$ is the Fermi temperature \cite{Zel2002physics}. The heat capacity of lattice in taken in Dulong-Petit limit, taking into consideration low Debye temperature of gold (around 170~K).

 The source term, appearing in Eq.~\ref{EqTTM_e} can be consequently describe by the following expression:

 \begin{equation}
 \label{EqTTM_S}
     S(t, z) = \sqrt{\frac{4\ln 2}{\pi}} \frac{\alpha (z)}{t_p} \phi \cdot e^{-4 \ln 2(t/t_p)^2},
 \end{equation}
 where $\phi$ is the laser fluence, $\alpha (z)$ is the spectraly integrated absorption coefficient as a function of depth in the material, and $t_p$ is the duration of the of the laser pulse taken at half maximum \cite{Carpene2006}. 

We have rigorously solved the set of coupled Eqs.~\ref{EqTTM_e} in the finite-volume PDE solver available in the FiPy package \cite{4814978}, and compared the results with the solution of "local$"$ TTM, that does not include diffusion term: 
 \begin{align}
     \label{EqTTM_le}
     C_{e0}T_e\frac{dT_e}{dt} &= G(T_l - T_e) + S(t),\\
     C_l \frac{dT_l}{dt} &= G(T_e - T_l) \label{EqTTM_ll},
 \end{align}
 The latter approach offers substantial ease of the computational complexity at the expense of ignoring the heat diffusion along the nanorods in the composite. While previous results have outlined the importance of accounting for the diffusion in some cases \cite{Nic2019}, our results suggest that, for excitation wavelength used in the modelling for this work (515~nm), the full solution is excessive and does not offer additional benefits. Due to this finding, we employed the model \ref{EqTTM_le}, solved independently for the nanorods and gold underlayer, to generate the transient spectra presented in this work.  

\subsection{Modelling transient optical response}\label{subsecOptM}

To model the transient permittivity of gold we employed a two-band model \cite{Guerrisi1975} and a procedure described in Ref.~\cite{Karaman2024}. Within this approach, the permittivity of gold $\epsilon_{Au} = \epsilon_{inter} + \epsilon_{intra}$ is separated into the single particle interband contribution, explicitly taking into account the optical transitions around X and L points of the band structure of gold, and a free-electron contribution, given by the Drude model:
\begin{align}
    \Im[\epsilon_{inter}(\hbar\omega,t)] = &\frac{A_X J_{X}(\hbar\omega,t)+A_{L_4^+} J_{L_4^+}(\hbar\omega,t)+A_{L_{5+6}^+} J_{L^+_{5+6}}(\hbar\omega,t)}{(\hbar\omega)^2} \\
    &\epsilon_{intra} =\epsilon_{\inf}-\frac{\omega_p^2}{\omega^2+i\omega\Gamma(\hbar\omega,T_e,T_l)}, 
\end{align}
where $\omega_p$ is the bulk plasma frequency, $A_i$ are the spectral weights of different interband transitions, and $J_{i}$ are the joint densities of states calculated by integrating the energy distribution of the joint density of states (EDJDOS) with the Fermi-Dirac occupation factors at a given electron temperature \cite{Masia2012,Guerrisi1975}. The damping $\Gamma(\hbar\omega,T_e,T_l) = \Gamma_{e-e}(\hbar\omega,T_e)+\Gamma_{e-ph}(T_l)$ contains the terms associated with the fractional Umklapp electron-electron and electron-phonon scatterings given by, respectively:
\begin{align}
    \Gamma_{e-e}(\hbar\omega,T_e) &=\frac{\pi^3\beta\Delta}{12\hbar E_F}\left[(k_BT_e)^2+(\hbar\omega/2\pi)^2\right] \\
    \Gamma_{e-ph}(T_l) &=\frac{1}{\tau_0}\bigg[\frac{2}{5}   +  4\bigg(\frac{T_l}{\Theta}\bigg)^5\int_{0}^{\Theta/T_l} \frac{z^4}{e^z-1}dz\bigg], 
\end{align}
where $\Delta$ = 0.75 and $\beta$ = 0.55 are dimensionless constants, characterising the electron-electron scattering, typical for noble metals \cite{Beach1977}, $\theta$ = 170~K is the Debye temperature, and $E_F$ = 5.53 eV is the Fermi energy.

The real part of the interband dielectric function of gold was obtained using the Kramers-Kronig transformation, and parameters $A_i$, $\epsilon_{inf}$ and $\tau_0$ were varied to match the tabulated values for the permittivity of gold at room temperature \cite{Johnson1972}.

Additionally, to represent the electrodeposited gold of the nanorods, that is known to demonstrate higher absorption than pure gold films due to polycrystallinity, we incorporate the "restricted$"$ mean free path $R$ for the free electrons into the dynamic model by augmenting the microscopic dielectric function of gold \cite{Johnson1972}:
\begin{equation}\label{eq_restrictedpath}
    \epsilon^{R}_{Au} = \epsilon_{Au} + \frac{i\omega_p^2\Gamma(L-R)}{\omega(\omega + i\Gamma)(\omega R + i\Gamma L)},
\end{equation}
where $L$ = 35 nm is the mean free path, corresponding to bulk gold. In accordance with the previous results on similar experimental samples, we set $R$ = 13 nm for the studied samples.

Transfer matrix model (TMM) was used to simulate the optical response of the metamaterial in the local effective medium approximation \cite{DJRoth2024}, which was checked to be valid for the metamaterial with the studied parameters for which the effects of spatial dispersion are negligible.

In order to qualitatively incorporate the effect of phonon vibrations in the simulated transient spectra, we used the numerical solutions of Eq.~\ref{vibrationModel}, parameterised with the experimental values for normal frequencies and acoustic damping. The breathing and extensional vibrations were assumed to contribute to the optical response by slightly modifying the filling factor and the rod length, respectively. Therefore, an additional term in the transient response was added according to
\begin{equation}
\frac{\Delta F}{F} = A\frac{\partial F(\lambda, r, l)}{\partial r} \Delta r(t) + B\frac{\partial F(\lambda, r, l)}{\partial l} \Delta l(t),
\end{equation}
where $F$ is either the reflection or transmission coefficient of the metamaterial, $\Delta r(t)$ and $\Delta l(t)$ are the time-dependent radius and length of nanorods, respectively, and A and B are fitting constants.


\bmhead{Acknowledgments}
This work was supported by the ERC iCOMM project (789340) and the UK EPSRC project EP/Y015673/1. The authors are grateful to H. Cossey for help with the cross-sectional SEM images. 
All the data supporting the findings of this work are presented in the results section and available from the corresponding author upon reasonable request.

\bibliography{sn-bibliography}


\begin{thebibliography}{43}
\ifx \bisbn   \undefined \def \bisbn  #1{ISBN #1}\fi
\ifx \binits  \undefined \def \binits#1{#1}\fi
\ifx \bauthor  \undefined \def \bauthor#1{#1}\fi
\ifx \batitle  \undefined \def \batitle#1{#1}\fi
\ifx \bjtitle  \undefined \def \bjtitle#1{#1}\fi
\ifx \bvolume  \undefined \def \bvolume#1{\textbf{#1}}\fi
\ifx \byear  \undefined \def \byear#1{#1}\fi
\ifx \bissue  \undefined \def \bissue#1{#1}\fi
\ifx \bfpage  \undefined \def \bfpage#1{#1}\fi
\ifx \blpage  \undefined \def \blpage #1{#1}\fi
\ifx \burl  \undefined \def \burl#1{\textsf{#1}}\fi
\ifx \doiurl  \undefined \def \doiurl#1{\url{https://doi.org/#1}}\fi
\ifx \betal  \undefined \def \betal{\textit{et al.}}\fi
\ifx \binstitute  \undefined \def \binstitute#1{#1}\fi
\ifx \binstitutionaled  \undefined \def \binstitutionaled#1{#1}\fi
\ifx \bctitle  \undefined \def \bctitle#1{#1}\fi
\ifx \beditor  \undefined \def \beditor#1{#1}\fi
\ifx \bpublisher  \undefined \def \bpublisher#1{#1}\fi
\ifx \bbtitle  \undefined \def \bbtitle#1{#1}\fi
\ifx \bedition  \undefined \def \bedition#1{#1}\fi
\ifx \bseriesno  \undefined \def \bseriesno#1{#1}\fi
\ifx \blocation  \undefined \def \blocation#1{#1}\fi
\ifx \bsertitle  \undefined \def \bsertitle#1{#1}\fi
\ifx \bsnm \undefined \def \bsnm#1{#1}\fi
\ifx \bsuffix \undefined \def \bsuffix#1{#1}\fi
\ifx \bparticle \undefined \def \bparticle#1{#1}\fi
\ifx \barticle \undefined \def \barticle#1{#1}\fi
\bibcommenthead
\ifx \bconfdate \undefined \def \bconfdate #1{#1}\fi
\ifx \botherref \undefined \def \botherref #1{#1}\fi
\ifx \url \undefined \def \url#1{\textsf{#1}}\fi
\ifx \bchapter \undefined \def \bchapter#1{#1}\fi
\ifx \bbook \undefined \def \bbook#1{#1}\fi
\ifx \bcomment \undefined \def \bcomment#1{#1}\fi
\ifx \oauthor \undefined \def \oauthor#1{#1}\fi
\ifx \citeauthoryear \undefined \def \citeauthoryear#1{#1}\fi
\ifx \endbibitem  \undefined \def \endbibitem {}\fi
\ifx \bconflocation  \undefined \def \bconflocation#1{#1}\fi
\ifx \arxivurl  \undefined \def \arxivurl#1{\textsf{#1}}\fi
\csname PreBibitemsHook\endcsname

\bibitem[\protect\citeauthoryear{Kauranen and Zayats}{2012}]{Kaur2012}
\begin{barticle}
\bauthor{\bsnm{Kauranen}, \binits{M.}},
\bauthor{\bsnm{Zayats}, \binits{A.V.}}:
\batitle{Nonlinear plasmonics}.
\bjtitle{Nature Photonics}
\bvolume{6}(\bissue{11}),
\bfpage{737}--\blpage{748}
(\byear{2012})
\end{barticle}
\endbibitem

\bibitem[\protect\citeauthoryear{Koya et~al.}{2023}]{Koya2023}
\begin{botherref}
\oauthor{\bsnm{Koya}, \binits{A.N.}},
\oauthor{\bsnm{Romanelli}, \binits{M.}},
\oauthor{\bsnm{Kuttruff}, \binits{J.}},
\oauthor{\bsnm{Henriksson}, \binits{N.}},
\oauthor{\bsnm{Stefancu}, \binits{A.}},
\oauthor{\bsnm{Grinblat}, \binits{G.}},
\oauthor{\bsnm{De~Andres}, \binits{A.}},
\oauthor{\bsnm{Schnur}, \binits{F.}},
\oauthor{\bsnm{Vanzan}, \binits{M.}},
\oauthor{\bsnm{Marsili}, \binits{M.}},
\oauthor{\bsnm{Rahaman}, \binits{M.}},
\oauthor{\bsnm{Viejo~Rodríguez}, \binits{A.}},
\oauthor{\bsnm{Tapani}, \binits{T.}},
\oauthor{\bsnm{Lin}, \binits{H.}},
\oauthor{\bsnm{Dana}, \binits{B.D.}},
\oauthor{\bsnm{Lin}, \binits{J.}},
\oauthor{\bsnm{Barbillon}, \binits{G.}},
\oauthor{\bsnm{Proietti~Zaccaria}, \binits{R.}},
\oauthor{\bsnm{Brida}, \binits{D.}},
\oauthor{\bsnm{Jariwala}, \binits{D.}},
\oauthor{\bsnm{Veisz}, \binits{L.}},
\oauthor{\bsnm{Cortés}, \binits{E.}},
\oauthor{\bsnm{Corni}, \binits{S.}},
\oauthor{\bsnm{Garoli}, \binits{D.}},
\oauthor{\bsnm{Maccaferri}, \binits{N.}}:
{Advances in ultrafast plasmonics}.
Applied Physics Reviews
\textbf{10}(2)
(2023)
\end{botherref}
\endbibitem

\bibitem[\protect\citeauthoryear{Khurgin et~al.}{2024}]{Khur2023}
\begin{barticle}
\bauthor{\bsnm{Khurgin}, \binits{J.}},
\bauthor{\bsnm{Bykov}, \binits{A.Y.}},
\bauthor{\bsnm{Zayats}, \binits{A.V.}}:
\batitle{Hot-electron dynamics in plasmonic nanostructures}.
\bjtitle{eLight}
\bvolume{4},
\bfpage{15}
(\byear{2024})
\end{barticle}
\endbibitem

\bibitem[\protect\citeauthoryear{Krasavin et~al.}{2018}]{Kra2018}
\begin{barticle}
\bauthor{\bsnm{Krasavin}, \binits{A.V.}},
\bauthor{\bsnm{Ginzburg}, \binits{P.}},
\bauthor{\bsnm{Zayats}, \binits{A.V.}}:
\batitle{Free-electron optical nonlinearities in plasmonic nanostructures: A review of the hydrodynamic description}.
\bjtitle{Laser \& Photonics Reviews}
\bvolume{12}(\bissue{1}),
\bfpage{1700082}
(\byear{2018})
\end{barticle}
\endbibitem

\bibitem[\protect\citeauthoryear{Del~Fatti et~al.}{2000}]{Fatti2000}
\begin{barticle}
\bauthor{\bsnm{Del~Fatti}, \binits{N.}},
\bauthor{\bsnm{Voisin}, \binits{C.}},
\bauthor{\bsnm{Achermann}, \binits{M.}},
\bauthor{\bsnm{Tzortzakis}, \binits{S.}},
\bauthor{\bsnm{Christofilos}, \binits{D.}},
\bauthor{\bsnm{Vall\'ee}, \binits{F.}}:
\batitle{Nonequilibrium electron dynamics in noble metals}.
\bjtitle{Phys. Rev. B}
\bvolume{61},
\bfpage{16956}--\blpage{16966}
(\byear{2000})
\end{barticle}
\endbibitem

\bibitem[\protect\citeauthoryear{Arbouet et~al.}{2003}]{Arbouet2003}
\begin{barticle}
\bauthor{\bsnm{Arbouet}, \binits{A.}},
\bauthor{\bsnm{Voisin}, \binits{C.}},
\bauthor{\bsnm{Christofilos}, \binits{D.}},
\bauthor{\bsnm{Langot}, \binits{P.}},
\bauthor{\bsnm{Fatti}, \binits{N.D.}},
\bauthor{\bsnm{Vall\'ee}, \binits{F.}},
\bauthor{\bsnm{Lerm\'e}, \binits{J.}},
\bauthor{\bsnm{Celep}, \binits{G.}},
\bauthor{\bsnm{Cottancin}, \binits{E.}},
\bauthor{\bsnm{Gaudry}, \binits{M.}},
\bauthor{\bsnm{Pellarin}, \binits{M.}},
\bauthor{\bsnm{Broyer}, \binits{M.}},
\bauthor{\bsnm{Maillard}, \binits{M.}},
\bauthor{\bsnm{Pileni}, \binits{M.P.}},
\bauthor{\bsnm{Treguer}, \binits{M.}}:
\batitle{Electron-phonon scattering in metal clusters}.
\bjtitle{Phys. Rev. Lett.}
\bvolume{90},
\bfpage{177401}
(\byear{2003})
\end{barticle}
\endbibitem

\bibitem[\protect\citeauthoryear{Nicholls et~al.}{2019}]{Nic2019}
\begin{barticle}
\bauthor{\bsnm{Nicholls}, \binits{L.H.}},
\bauthor{\bsnm{Stefaniuk}, \binits{T.}},
\bauthor{\bsnm{Nasir}, \binits{M.E.}},
\bauthor{\bsnm{Rodr{\'i}guez-Fortu{\~{n}}o}, \binits{F.J.}},
\bauthor{\bsnm{Wurtz}, \binits{G.A.}},
\bauthor{\bsnm{Zayats}, \binits{A.V.}}:
\batitle{Designer photonic dynamics by using non-uniform electron temperature distribution for on-demand all-optical switching times}.
\bjtitle{Nature Communications}
\bvolume{10}(\bissue{1}),
\bfpage{2967}
(\byear{2019})
\end{barticle}
\endbibitem

\bibitem[\protect\citeauthoryear{Temnov}{2012}]{Temnov2012}
\begin{barticle}
\bauthor{\bsnm{Temnov}, \binits{V.V.}}:
\batitle{Ultrafast acousto-magneto-plasmonics}.
\bjtitle{Nature Photonics}
\bvolume{6}(\bissue{11}),
\bfpage{728}--\blpage{736}
(\byear{2012})
\end{barticle}
\endbibitem

\bibitem[\protect\citeauthoryear{Frischwasser et~al.}{2022}]{Frischwasser2022}
\begin{barticle}
\bauthor{\bsnm{Frischwasser}, \binits{K.}},
\bauthor{\bsnm{Cohen}, \binits{K.}},
\bauthor{\bsnm{Tsesses}, \binits{S.}},
\bauthor{\bsnm{Dolev}, \binits{S.}},
\bauthor{\bsnm{Rosenblatt}, \binits{G.}},
\bauthor{\bsnm{Bartal}, \binits{G.}}:
\batitle{Nonlinear forced response of plasmonic nanostructures}.
\bjtitle{Phys. Rev. Lett.}
\bvolume{128},
\bfpage{103901}
(\byear{2022})
\end{barticle}
\endbibitem

\bibitem[\protect\citeauthoryear{Pelton et~al.}{2009}]{Pelton2009}
\begin{barticle}
\bauthor{\bsnm{Pelton}, \binits{M.}},
\bauthor{\bsnm{Sader}, \binits{J.E.}},
\bauthor{\bsnm{Burgin}, \binits{J.}},
\bauthor{\bsnm{Liu}, \binits{M.}},
\bauthor{\bsnm{Guyot-Sionnest}, \binits{P.}},
\bauthor{\bsnm{Gosztola}, \binits{D.}}:
\batitle{Damping of acoustic vibrations in gold nanoparticles}.
\bjtitle{Nature Nanotechnology}
\bvolume{4}(\bissue{8}),
\bfpage{492}--\blpage{495}
(\byear{2009})
\end{barticle}
\endbibitem

\bibitem[\protect\citeauthoryear{Bykov et~al.}{2023}]{Bykov2021}
\begin{barticle}
\bauthor{\bsnm{Bykov}, \binits{A.Y.}},
\bauthor{\bsnm{Xie}, \binits{Y.X.}},
\bauthor{\bsnm{Krasavin}, \binits{A.V.A.V.Z.}},
\bauthor{\bsnm{Zayats}, \binits{A.V.}}:
\batitle{Broadband transient response and wavelength-tunable photoacoustics in plasmonic hetero-nanoparticles}.
\bjtitle{Nano Letters}
\bvolume{23}(\bissue{7}),
\bfpage{2786}--\blpage{2791}
(\byear{2023})
\end{barticle}
\endbibitem

\bibitem[\protect\citeauthoryear{Ruello et~al.}{2015}]{Ruello2015}
\begin{barticle}
\bauthor{\bsnm{Ruello}, \binits{P.}},
\bauthor{\bsnm{Ayouch}, \binits{A.}},
\bauthor{\bsnm{Vaudel}, \binits{G.}},
\bauthor{\bsnm{Pezeril}, \binits{T.}},
\bauthor{\bsnm{Delorme}, \binits{N.}},
\bauthor{\bsnm{Sato}, \binits{S.}},
\bauthor{\bsnm{Kimura}, \binits{K.}},
\bauthor{\bsnm{Gusev}, \binits{V.E.}}:
\batitle{Ultrafast acousto-plasmonics in gold nanoparticle superlattices}.
\bjtitle{Phys. Rev. B}
\bvolume{92},
\bfpage{174304}
(\byear{2015})
\end{barticle}
\endbibitem

\bibitem[\protect\citeauthoryear{Guyader et~al.}{2008}]{LeGuyader_2008}
\begin{barticle}
\bauthor{\bsnm{Guyader}, \binits{L.L.}},
\bauthor{\bsnm{Kirilyuk}, \binits{A.}},
\bauthor{\bsnm{Rasing}, \binits{T.}},
\bauthor{\bsnm{Wurtz}, \binits{G.A.}},
\bauthor{\bsnm{Zayats}, \binits{A.V.}},
\bauthor{\bsnm{Alkemade}, \binits{P.F.A.}},
\bauthor{\bsnm{Smolyaninov}, \binits{I.I.}}:
\batitle{Coherent control of surface plasmon polariton mediated optical transmission}.
\bjtitle{Journal of Physics D: Applied Physics}
\bvolume{41}(\bissue{19}),
\bfpage{195102}
(\byear{2008})
\end{barticle}
\endbibitem

\bibitem[\protect\citeauthoryear{Bragas et~al.}{2023}]{Bragas2023}
\begin{barticle}
\bauthor{\bsnm{Bragas}, \binits{A.V.}},
\bauthor{\bsnm{Maier}, \binits{S.A.}},
\bauthor{\bsnm{Boggiano}, \binits{H.D.}},
\bauthor{\bsnm{Grinblat}, \binits{G.}},
\bauthor{\bsnm{Bert\'{e}}, \binits{R.}},
\bauthor{\bsnm{S.~Menezes}, \binits{L.}},
\bauthor{\bsnm{Cort\'{e}s}, \binits{E.}}:
\batitle{Nanomechanics with plasmonic nanoantennas: ultrafast and local exchange between electromagnetic and mechanical energy}.
\bjtitle{J. Opt. Soc. Am. B}
\bvolume{40}(\bissue{5}),
\bfpage{1196}--\blpage{1211}
(\byear{2023})
\end{barticle}
\endbibitem

\bibitem[\protect\citeauthoryear{Yu and Ouyang}{2018}]{YuOuyang2018}
\begin{barticle}
\bauthor{\bsnm{Yu}, \binits{S.-J.}},
\bauthor{\bsnm{Ouyang}, \binits{M.}}:
\batitle{Coherent discriminatory modal manipulation of acoustic phonons at the nanoscale}.
\bjtitle{Nano Letters}
\bvolume{18}(\bissue{2}),
\bfpage{1124}--\blpage{1129}
(\byear{2018})
\end{barticle}
\endbibitem

\bibitem[\protect\citeauthoryear{Wurtz et~al.}{2011}]{Wurtz2011}
\begin{barticle}
\bauthor{\bsnm{Wurtz}, \binits{G.A.}},
\bauthor{\bsnm{Pollard}, \binits{R.}},
\bauthor{\bsnm{Hendren}, \binits{W.}},
\bauthor{\bsnm{Wiederrecht}, \binits{G.P.}},
\bauthor{\bsnm{Gosztola}, \binits{D.J.}},
\bauthor{\bsnm{Podolskiy}, \binits{V.A.}},
\bauthor{\bsnm{Zayats}, \binits{A.V.}}:
\batitle{Designed ultrafast optical nonlinearity in a plasmonic nanorod metamaterial enhanced by nonlocality}.
\bjtitle{Nature Nanotechnology}
\bvolume{6}(\bissue{2}),
\bfpage{107}--\blpage{111}
(\byear{2011})
\end{barticle}
\endbibitem

\bibitem[\protect\citeauthoryear{Poddubny et~al.}{2013}]{Poddubny2013}
\begin{barticle}
\bauthor{\bsnm{Poddubny}, \binits{A.}},
\bauthor{\bsnm{Iorsh}, \binits{I.}},
\bauthor{\bsnm{Belov}, \binits{P.}},
\bauthor{\bsnm{Kivshar}, \binits{Y.}}:
\batitle{Hyperbolic metamaterials}.
\bjtitle{Nature Photonics}
\bvolume{7}(\bissue{12}),
\bfpage{948}--\blpage{957}
(\byear{2013})
\end{barticle}
\endbibitem

\bibitem[\protect\citeauthoryear{Roth et~al.}{2024}]{DJRoth2024}
\begin{barticle}
\bauthor{\bsnm{Roth}, \binits{D.J.}},
\bauthor{\bsnm{Krasavin}, \binits{A.V.}},
\bauthor{\bsnm{Zayats}, \binits{A.V.}}:
\batitle{Nanophotonics with plasmonic nanorod metamaterials}.
\bjtitle{Laser \& Photonics Reviews}
\bvolume{18}(\bissue{8}),
\bfpage{2300886}
(\byear{2024})
\end{barticle}
\endbibitem

\bibitem[\protect\citeauthoryear{Reshef et~al.}{2019}]{Reshef2019}
\begin{barticle}
\bauthor{\bsnm{Reshef}, \binits{O.}},
\bauthor{\bsnm{De~Leon}, \binits{I.}},
\bauthor{\bsnm{Alam}, \binits{M.Z.}},
\bauthor{\bsnm{Boyd}, \binits{R.W.}}:
\batitle{Nonlinear optical effects in epsilon-near-zero media}.
\bjtitle{Nature Reviews Materials}
\bvolume{4}(\bissue{8}),
\bfpage{535}--\blpage{551}
(\byear{2019})
\end{barticle}
\endbibitem

\bibitem[\protect\citeauthoryear{Neira et~al.}{2015}]{Neira2015}
\begin{barticle}
\bauthor{\bsnm{Neira}, \binits{A.D.}},
\bauthor{\bsnm{Olivier}, \binits{N.}},
\bauthor{\bsnm{Nasir}, \binits{M.E.}},
\bauthor{\bsnm{Dickson}, \binits{W.}},
\bauthor{\bsnm{Wurtz}, \binits{G.A.}},
\bauthor{\bsnm{Zayats}, \binits{A.V.}}:
\batitle{Eliminating material constraints for nonlinearity with plasmonic metamaterials}.
\bjtitle{Nature Communications}
\bvolume{6}(\bissue{1}),
\bfpage{7757}
(\byear{2015})
\end{barticle}
\endbibitem

\bibitem[\protect\citeauthoryear{Vasilantonakis et~al.}{2015}]{Vasilantonakis2015}
\begin{barticle}
\bauthor{\bsnm{Vasilantonakis}, \binits{N.}},
\bauthor{\bsnm{Nasir}, \binits{M.E.}},
\bauthor{\bsnm{Dickson}, \binits{W.}},
\bauthor{\bsnm{Wurtz}, \binits{G.A.}},
\bauthor{\bsnm{Zayats}, \binits{A.V.}}:
\batitle{Bulk plasmon-polaritons in hyperbolic nanorod metamaterial waveguides}.
\bjtitle{Laser \& Photonics Reviews}
\bvolume{9}(\bissue{3}),
\bfpage{345}--\blpage{353}
(\byear{2015})
\end{barticle}
\endbibitem

\bibitem[\protect\citeauthoryear{Schirato et~al.}{2020}]{Schirato2020}
\begin{barticle}
\bauthor{\bsnm{Schirato}, \binits{A.}},
\bauthor{\bsnm{Maiuri}, \binits{M.}},
\bauthor{\bsnm{Toma}, \binits{A.}},
\bauthor{\bsnm{Fugattini}, \binits{S.}},
\bauthor{\bsnm{Proietti~Zaccaria}, \binits{R.}},
\bauthor{\bsnm{Laporta}, \binits{P.}},
\bauthor{\bsnm{Nordlander}, \binits{P.}},
\bauthor{\bsnm{Cerullo}, \binits{G.}},
\bauthor{\bsnm{Alabastri}, \binits{A.}},
\bauthor{\bsnm{Della~Valle}, \binits{G.}}:
\batitle{Transient optical symmetry breaking for ultrafast broadband dichroism in plasmonic metasurfaces}.
\bjtitle{Nature Photonics}
\bvolume{14}(\bissue{12}),
\bfpage{723}--\blpage{727}
(\byear{2020})
\end{barticle}
\endbibitem

\bibitem[\protect\citeauthoryear{Yu et~al.}{2013}]{Yu2013}
\begin{barticle}
\bauthor{\bsnm{Yu}, \binits{K.}},
\bauthor{\bsnm{Zijlstra}, \binits{P.}},
\bauthor{\bsnm{Sader}, \binits{J.E.}},
\bauthor{\bsnm{Xu}, \binits{Q.-H.}},
\bauthor{\bsnm{Orrit}, \binits{M.}}:
\batitle{Damping of acoustic vibrations of immobilized single gold nanorods in different environments}.
\bjtitle{Nano Letters}
\bvolume{13}(\bissue{6}),
\bfpage{2710}--\blpage{2716}
(\byear{2013})
\end{barticle}
\endbibitem

\bibitem[\protect\citeauthoryear{Wang and Guo}{2007}]{Wang2007}
\begin{barticle}
\bauthor{\bsnm{Wang}, \binits{J.}},
\bauthor{\bsnm{Guo}, \binits{C.}}:
\batitle{Effect of electron heating on femtosecond laser-induced coherent acoustic phonons in noble metals}.
\bjtitle{Phys. Rev. B}
\bvolume{75},
\bfpage{184304}
(\byear{2007})
\end{barticle}
\endbibitem

\bibitem[\protect\citeauthoryear{Wang et~al.}{2007}]{WangJ2007}
\begin{barticle}
\bauthor{\bsnm{Wang}, \binits{J.}},
\bauthor{\bsnm{Wu}, \binits{J.}},
\bauthor{\bsnm{Guo}, \binits{C.}}:
\batitle{Resolving dynamics of acoustic phonons by surface plasmons}.
\bjtitle{Opt. Lett.}
\bvolume{32}(\bissue{6}),
\bfpage{719}--\blpage{721}
(\byear{2007})
\end{barticle}
\endbibitem

\bibitem[\protect\citeauthoryear{Perner et~al.}{2000}]{Perner2000}
\begin{barticle}
\bauthor{\bsnm{Perner}, \binits{M.}},
\bauthor{\bsnm{Gresillon}, \binits{S.}},
\bauthor{\bsnm{M\"arz}, \binits{J.}},
\bauthor{\bsnm{Plessen}, \binits{G.}},
\bauthor{\bsnm{Feldmann}, \binits{J.}},
\bauthor{\bsnm{Porstendorfer}, \binits{J.}},
\bauthor{\bsnm{Berg}, \binits{K.-J.}},
\bauthor{\bsnm{Berg}, \binits{G.}}:
\batitle{Observation of hot-electron pressure in the vibration dynamics of metal nanoparticles}.
\bjtitle{Phys. Rev. Lett.}
\bvolume{85},
\bfpage{792}--\blpage{795}
(\byear{2000})
\end{barticle}
\endbibitem

\bibitem[\protect\citeauthoryear{Tas and Maris}{1994}]{PhysRevB.49.15046}
\begin{barticle}
\bauthor{\bsnm{Tas}, \binits{G.}},
\bauthor{\bsnm{Maris}, \binits{H.J.}}:
\batitle{Electron diffusion in metals studied by picosecond ultrasonics}.
\bjtitle{Phys. Rev. B}
\bvolume{49},
\bfpage{15046}--\blpage{15054}
(\byear{1994})
\end{barticle}
\endbibitem

\bibitem[\protect\citeauthoryear{Gusev}{1992}]{GUSEV199276}
\begin{barticle}
\bauthor{\bsnm{Gusev}, \binits{V.E.}}:
\batitle{On the duration of acoustic pulses excited by subpicosecond laser action on metals}.
\bjtitle{Optics Communications}
\bvolume{94}(\bissue{1}),
\bfpage{76}--\blpage{78}
(\byear{1992})
\end{barticle}
\endbibitem

\bibitem[\protect\citeauthoryear{Thomsen et~al.}{1986}]{Thomsen1986}
\begin{barticle}
\bauthor{\bsnm{Thomsen}, \binits{C.}},
\bauthor{\bsnm{Grahn}, \binits{H.T.}},
\bauthor{\bsnm{Maris}, \binits{H.J.}},
\bauthor{\bsnm{Tauc}, \binits{J.}}:
\batitle{Surface generation and detection of phonons by picosecond light pulses}.
\bjtitle{Phys. Rev. B}
\bvolume{34},
\bfpage{4129}--\blpage{4138}
(\byear{1986})
\end{barticle}
\endbibitem

\bibitem[\protect\citeauthoryear{Deacon et~al.}{2017}]{Deacon2017}
\begin{barticle}
\bauthor{\bsnm{Deacon}, \binits{W.M.}},
\bauthor{\bsnm{Lombardi}, \binits{A.}},
\bauthor{\bsnm{Benz}, \binits{F.}},
\bauthor{\bsnm{Valle-Inclan~Redondo}, \binits{Y.}},
\bauthor{\bsnm{Chikkaraddy}, \binits{R.}},
\bauthor{\bsnm{Nijs}, \binits{B.}},
\bauthor{\bsnm{Kleemann}, \binits{M.-E.}},
\bauthor{\bsnm{Mertens}, \binits{J.}},
\bauthor{\bsnm{Baumberg}, \binits{J.J.}}:
\batitle{Interrogating nanojunctions using ultraconfined acoustoplasmonic coupling}.
\bjtitle{Phys. Rev. Lett.}
\bvolume{119},
\bfpage{023901}
(\byear{2017})
\end{barticle}
\endbibitem

\bibitem[\protect\citeauthoryear{Frenzel et~al.}{2019}]{Frenzel2019}
\begin{barticle}
\bauthor{\bsnm{Frenzel}, \binits{T.}},
\bauthor{\bsnm{K{\"o}pfler}, \binits{J.}},
\bauthor{\bsnm{Jung}, \binits{E.}},
\bauthor{\bsnm{Kadic}, \binits{M.}},
\bauthor{\bsnm{Wegener}, \binits{M.}}:
\batitle{Ultrasound experiments on acoustical activity in chiral mechanical metamaterials}.
\bjtitle{Nature Communications}
\bvolume{10}(\bissue{1}),
\bfpage{3384}
(\byear{2019})
\end{barticle}
\endbibitem

\bibitem[\protect\citeauthoryear{Jansen et~al.}{2023}]{Jansen2023}
\begin{barticle}
\bauthor{\bsnm{Jansen}, \binits{M.}},
\bauthor{\bsnm{Tisdale}, \binits{W.A.}},
\bauthor{\bsnm{Wood}, \binits{V.}}:
\batitle{Nanocrystal phononics}.
\bjtitle{Nature Materials}
\bvolume{22}(\bissue{2}),
\bfpage{161}--\blpage{169}
(\byear{2023})
\end{barticle}
\endbibitem

\bibitem[\protect\citeauthoryear{Zaleska et~al.}{2024}]{Zaleska2024}
\begin{barticle}
\bauthor{\bsnm{Zaleska}, \binits{A.}},
\bauthor{\bsnm{Krasavin}, \binits{A.V.}},
\bauthor{\bsnm{Zayats}, \binits{A.V.}},
\bauthor{\bsnm{Dickson}, \binits{W.}}:
\batitle{Copper-based core–shell metamaterials with ultra-broadband and reversible enz tunability}.
\bjtitle{Mater. Adv.}
\bvolume{5},
\bfpage{5845}--\blpage{5854}
(\byear{2024})
\end{barticle}
\endbibitem

\bibitem[\protect\citeauthoryear{Peruch et~al.}{2017}]{Peruch2017}
\begin{barticle}
\bauthor{\bsnm{Peruch}, \binits{S.}},
\bauthor{\bsnm{Neira}, \binits{A.}},
\bauthor{\bsnm{Wurtz}, \binits{G.A.}},
\bauthor{\bsnm{Wells}, \binits{B.}},
\bauthor{\bsnm{Podolskiy}, \binits{V.A.}},
\bauthor{\bsnm{Zayats}, \binits{A.V.}}:
\batitle{Geometry defines ultrafast hot-carrier dynamics and kerr nonlinearity in plasmonic metamaterial waveguides and cavities}.
\bjtitle{Advanced Optical Materials}
\bvolume{5}(\bissue{15}),
\bfpage{1700299}
(\byear{2017})
\end{barticle}
\endbibitem

\bibitem[\protect\citeauthoryear{Chen et~al.}{2006}]{Chen2006307}
\begin{barticle}
\bauthor{\bsnm{Chen}, \binits{J.K.}},
\bauthor{\bsnm{Tzou}, \binits{D.Y.}},
\bauthor{\bsnm{Beraun}, \binits{J.E.}}:
\batitle{A semiclassical two-temperature model for ultrafast laser heating}.
\bjtitle{International Journal of Heat and Mass Transfer}
\bvolume{49}(\bissue{1}),
\bfpage{307}--\blpage{316}
(\byear{2006})
\end{barticle}
\endbibitem

\bibitem[\protect\citeauthoryear{Zel'dovich and Raizer}{2002}]{Zel2002physics}
\begin{bbook}
\bauthor{\bsnm{Zel'dovich}, \binits{Y.B.}},
\bauthor{\bsnm{Raizer}, \binits{Y.P.}}:
\bbtitle{Physics of Shock Waves and High-Temperature Hydrodynamic Phenomena}.
\bpublisher{Dover Publications},
\blocation{New {Y}ork}
(\byear{2002})
\end{bbook}
\endbibitem

\bibitem[\protect\citeauthoryear{Carpene}{2006}]{Carpene2006}
\begin{barticle}
\bauthor{\bsnm{Carpene}, \binits{E.}}:
\batitle{Ultrafast laser irradiation of metals: Beyond the two-temperature model}.
\bjtitle{Phys. Rev. B}
\bvolume{74},
\bfpage{024301}
(\byear{2006})
\end{barticle}
\endbibitem

\bibitem[\protect\citeauthoryear{Guyer et~al.}{2009}]{4814978}
\begin{barticle}
\bauthor{\bsnm{Guyer}, \binits{J.E.}},
\bauthor{\bsnm{Wheeler}, \binits{D.}},
\bauthor{\bsnm{Warren}, \binits{J.A.}}:
\batitle{Fipy: Partial differential equations with python}.
\bjtitle{Computing in Science \& Engineering}
\bvolume{11}(\bissue{3}),
\bfpage{6}--\blpage{15}
(\byear{2009})
\end{barticle}
\endbibitem

\bibitem[\protect\citeauthoryear{Guerrisi et~al.}{1975}]{Guerrisi1975}
\begin{barticle}
\bauthor{\bsnm{Guerrisi}, \binits{M.}},
\bauthor{\bsnm{Rosei}, \binits{R.}},
\bauthor{\bsnm{Winsemius}, \binits{P.}}:
\batitle{Splitting of the interband absorption edge in au}.
\bjtitle{Phys. Rev. B}
\bvolume{12},
\bfpage{557}--\blpage{563}
(\byear{1975})
\end{barticle}
\endbibitem

\bibitem[\protect\citeauthoryear{Karaman et~al.}{2024}]{Karaman2024}
\begin{barticle}
\bauthor{\bsnm{Karaman}, \binits{C.O.}},
\bauthor{\bsnm{Bykov}, \binits{A.Y.}},
\bauthor{\bsnm{Kiani}, \binits{F.}},
\bauthor{\bsnm{Tagliabue}, \binits{G.}},
\bauthor{\bsnm{Zayats}, \binits{A.V.}}:
\batitle{Ultrafast hot-carrier dynamics in ultrathin monocrystalline gold}.
\bjtitle{Nature Communications}
\bvolume{15}(\bissue{1}),
\bfpage{703}
(\byear{2024})
\end{barticle}
\endbibitem

\bibitem[\protect\citeauthoryear{Masia et~al.}{2012}]{Masia2012}
\begin{barticle}
\bauthor{\bsnm{Masia}, \binits{F.}},
\bauthor{\bsnm{Langbein}, \binits{W.}},
\bauthor{\bsnm{Borri}, \binits{P.}}:
\batitle{Measurement of the dynamics of plasmons inside individual gold nanoparticles using a femtosecond phase-resolved microscope}.
\bjtitle{Phys. Rev. B}
\bvolume{85},
\bfpage{235403}
(\byear{2012})
\end{barticle}
\endbibitem

\bibitem[\protect\citeauthoryear{Beach and Christy}{1977}]{Beach1977}
\begin{barticle}
\bauthor{\bsnm{Beach}, \binits{R.T.}},
\bauthor{\bsnm{Christy}, \binits{R.W.}}:
\batitle{Electron-electron scattering in the intraband optical conductivity of cu, ag, and au}.
\bjtitle{Phys. Rev. B}
\bvolume{16},
\bfpage{5277}--\blpage{5284}
(\byear{1977})
\end{barticle}
\endbibitem

\bibitem[\protect\citeauthoryear{Johnson and Christy}{1972}]{Johnson1972}
\begin{barticle}
\bauthor{\bsnm{Johnson}, \binits{P.B.}},
\bauthor{\bsnm{Christy}, \binits{R.W.}}:
\batitle{Optical constants of the noble metals}.
\bjtitle{Phys. Rev. B}
\bvolume{6},
\bfpage{4370}--\blpage{4379}
(\byear{1972})
\end{barticle}
\endbibitem

\end{thebibliography}

\end{document}